\documentclass{article}

\usepackage{arxiv}

\usepackage[utf8]{inputenc} 
\usepackage[T1]{fontenc}    
\usepackage{hyperref}       
\usepackage{url}            
\usepackage{booktabs}       
\usepackage{amsfonts}       
\usepackage{nicefrac}       
\usepackage{microtype}      
\usepackage{lipsum}
\usepackage{graphicx}
\graphicspath{ {./images/} }

\usepackage{graphicx}
\usepackage{dcolumn}
\usepackage{bm}
\usepackage{comment}

\usepackage{graphicx}
\usepackage{epsfig}
\usepackage{dcolumn}
\usepackage[up]{subfigure}
\usepackage{caption}

\usepackage{float}
\usepackage{dsfont}	
\usepackage{relsize}
\usepackage{verbatim}

\usepackage{epstopdf}

\usepackage{amsfonts}
\newcommand\norm[1]{\left\lVert#1\right\rVert}
\usepackage{float}
\usepackage{ragged2e}
\usepackage{braket}
\usepackage[font=small,labelfont=bf,justification=raggedright, format=plain]{caption}
\usepackage{setspace}
\usepackage{amsmath}

\title{One-Dimensional Lazy Quantum Walk in Ternary System}

\author{
 Amit Saha \\
  A. K. Choudhury School of Information Technology \\
   University of Calcutta\\
  Kolkata \\
  \texttt{abamitsaha@gmail.com} \\
   \And
 Sudhindu Bikash Mandal \\
  Regent institute of science and technology, \\
  India\\
  \texttt{sudhindu.mandal@gmail.com } \\
  \And
 Debasri Saha \\
  A. K. Choudhury School of Information Technology \\
   University of Calcutta\\
  Kolkata \\
  \texttt{debasri\_cu@yahoo.in} \\
  \AND
  Amlan Chakrabarti \\
  A. K. Choudhury School of Information Technology \\
   University of Calcutta\\
  Kolkata \\
  \texttt{acakcs@caluniv.ac.in}
}

\begin{document}
\maketitle
\begin{abstract}
Quantum walks play an important role for developing quantum algorithms and quantum simulations. Here we present  one dimensional three-state quantum walk(lazy quantum walk) and show its equivalence for circuit realization in ternary quantum logic for the first of its kind. Using an appropriate logical mapping of the position space on which a walker evolves onto the multi-qutrit states, we present efficient quantum circuits considering the nearest neighbour position space for the implementation of lazy quantum walks in one-dimensional position space in ternary quantum system. We also address scalability in terms of $n$-qutrit ternary system with example circuits for a three qutrit state space.

\end{abstract}
\maketitle
%
%
%
%
%

\section{Introduction}

As development of quantum computers has achieved a remarkable success in recent years, everyone has shown a great interest to implement quantum algorithms, which give a potential speedup over their classical counterparts \cite{nielsen_chuang_2010, Farhi_1998}. Quantum walk \cite{PhysRevA.48.1687} is one such quantum algorithm that can be a prospective candidate for solving search problems \cite{Shenvi_2003} with numerous speedup over the conventional computer. 

For the last 25 years, researchers have been sincerely working on quantum walks and its applications \cite{ambainis2004quantum, aharonov2000quantum, magniez2003quantum}. Quantum walks has two main variants, one is discrete-time quantum walk (DTQW) \cite{ambainis2003quantum, Tregenna_2003} and another is continuous-time quantum walk (CTQW) \cite{Childs_2003}. The discrete-time quantum walk (DTQW) is defined on the combination of coin (particle) and position Hilbert space. The evolution of this position space is driven by a position shift operator controlled by a quantum coin operator. The continuous-time quantum walk (CTQW) is defined only on the position Hilbert space. In this variant, the evolution of this position space is driven by the Hamiltonian of the system.. The probability distribution of particles for both the variants of the quantum walks spreads quadratically faster in position space compared to the classical random walk \cite{Kempe03, Venegas_Andraca_2012, Kendon_2006}, in DTQW two-state Hadamard coin operator has been used. A concept of lazy quantum walk(LQW) \cite{Childs_2009} was later introduced incorporating three-state quantum coin operator. This helps to establish a relationship between CTQW and DTQW. Further behavioral analysis of DTQW has been carried out by making use of three state quantum coin on line and cycle \cite{Falkner_2014, machida2014limit, endo2019stationary, kajiwara2019periodicity}. Soon it was shown in \cite{Li_2015} that the occupancy rate of lazy quantum walk along one-dimensional line is better than DTQW . A small variation in lazy quantum walk gave birth to a breakthrough algorithm, Lackadaisical quantum walk \cite{Wong_2015} which gives algorithmic speedup than the previous ones \cite{Wong_2018, saha2018search}.


Due to developments of quantum computers in the past two-three years by the organizations like Google, IBM, Microsoft etc., implementation of quantum circuits with small number of qubits is now at our fingertips. 
In 2017, IBM unveiled their first 5-qubit quantum processor 'IBM-Q' and immediately after that physical realization of topological quantum walks on IBM-Q was carried out \cite{balu2017physical}. In \cite{Yan753}, a strongly correlated quantum walks has been implemented on 12-qubit superconducting processor in 2019. Quantum walk has also been implemented on trapped-ion based quantum computer in 2020 \cite{alderete2020quantum}. After that implementation of discrete-time quantum walks with one and two interacting walkers on the cycle and the two-dimensional lattice on IBM quantum computers have been carried out in \cite{acasiete2020experimental}. In \cite{singh2020universal}, Authors proposed an efficient quantum circuits realization to implement DTQWs in one-dimensional position space on a five-qubit processor. These circuits can be implemented on any of the present superconducting qubit, trapped ion qubit or other circuit based quantum devices. They further showed that these circuits can be scaled up to implement more steps and to higher spatial dimensions, generalized to implement multi-particle DTQWs, and DTQW based algorithms. 

But implementation of discrete-time quantum walk (DTQW) with superconducting qubits is difficult since on-chip superconducting qubits cannot hop between lattice sites. To overcome this problem, an efficient protocol has been proposed for the implementation of DTQW in circuit quantum electrodynamics (QED), in which only $N+1$ qutrits and $N$ assistant cavities are needed for an $N$-step DTQW \cite{Zhou_2019}. As qutrits can be used to encode more information, more researchers are fascinated towards working with ternary quantum system in recent years. For a ternary quantum system, the unit of information is known as a $qutrit$ and the corresponding quantum system can be defined using the orthonormal basis states $\ket{0}$, $\ket{1}$, $\ket{2}$ \cite{Muthukrishnan_2000}.

Of late several works have been done on quantum algorithms in ternary quantum system \cite{Fan_2007, bocharov2015improved, Bocharov_2017}. Search problems have also been dealt with the help of well-known quantum search algorithms \cite{6845018, Ivanov_2012}, which have been implemented using ternary elementary gates \cite{Muthukrishnan_2000, doi:10.1142/S0218126615501212, di2011elementary, Di_2013}. In recent past, researchers have claimed to develop a superconducting qutrit processor \cite{blok2020quantum}. To add to that researchers have also claimed to have implemented Walsh-Hadamard gate on this superconducting qutrit processor \cite{yurtalan2020implementation}. This gives a ray of hope that ternary quantum computers will soon come into play. This led us to carry out the circuit realization of one-dimensional lazy quantum walk in ternary system in this paper. Such that whenever ternary quantum computers are live in action we can straightaway map them to it to get the advantage in application of lazy quantum walk.

In this paper, we have implemented one-dimensional quantum walk using three state coin in ternary quantum system. Our novelty lies in the fact that:

\begin{itemize}
    \item We define one dimensional discrete-time quantum walk using three-state coin(LQW) in ternary quantum system. 
    
    \item Further, we propose an efficient quantum circuit realization to implement one dimensional discrete-time quantum walk using three-state coin(LQW) in ternary quantum system for the first of its kind using an appropriate logical mapping of the position space considering nearest neighbour logic on which a walker evolves onto the multi-qutrit states.
     
    \item We also address scalability of the proposed circuit in terms of $n$-qutrit system, which makes this circuit realization generalized in nature for implementing more steps.
\end{itemize}

The structure of this paper is as follows. Section. II describes the dynamics of one-dimensional quantum walk in binary quantum system. Section. III defines the dynamics of one-dimensional quantum walk using three-state quantum coin(LQW) in ternary system. Section. IV proposes efficient quantum circuit implementation for one-dimensional quantum walk using three-state quantum coin(LQW) in ternary system followed by the generalization of quantum circuit. Section. V describes our conclusions.

\section{One-Dimensional Discrete Time Quantum Walk}

Discrete time quantum walks (DTQW) take place in the product space $\mathcal{H}={\mathcal{H}}_{p}\bigotimes {\mathcal{H}}_{c}$. ${\mathcal{H}}_{p}$ is a Hilbert space which has orthonormal basis given by the position states $\{|x\rangle, x\in\mathcal{Z}\}$. The default initial position state is $|0\rangle$. Due to the two choices of the movement, DTQW have a two-dimensional coin. Therefore, $\mathcal{H}_{c}$ is a Hilbert space spanned by the orthonormal basis$\{\ket{\uparrow},  \ket{\downarrow}\}$ ($\uparrow$ for right and $\downarrow$ for left).

Let $|x,\alpha\rangle$ be a basis state, where $x\in \mathcal{Z}$ represents the position of the particle and $\alpha\in \{\uparrow,  \downarrow\}$ represents the coin state. The evolution of the whole system at each step of the walk can be described by the unitary operator denoted by $U$,
\begin{equation}\label{21}
    U=S({\mathcal{I}}\otimes C),
\end{equation}where $S$ is the shift operator defined by
\begin{equation} \label{Shift}
S  = \sum_{x\in\mathbb{Z}} \bigg (\ket{\uparrow}\bra{\uparrow}
\otimes   \ket{x-1}\bra{x}+\ket{\downarrow}\bra{\downarrow} \otimes \ket{x+1}\bra{x}\bigg ).
\end{equation}

$I$ is the identity matrix which operates in $\mathcal{H}_{p}$, while $C$ is the coin operation. Hadamard coin is an example of two-dimensional quantum coin operator, which is denoted by $H$,
\begin{equation}
    C=\frac{1}{\sqrt{2}}\left(
          \begin{array}{ccc}
            1 & 1 \\
            1 & -1 \\
          \end{array}
        \right).
\end{equation}

\begin{figure}[h]
\centering
\includegraphics[width=7cm, scale=1]{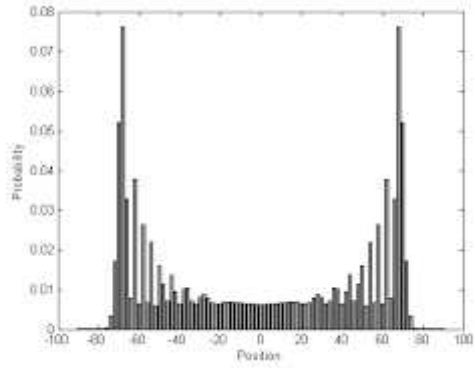}
\captionof{figure}{Quantum walks on a line using Hadamard coin in binary system, after 100 steps}
\label{fig1}
\end{figure}

The state of the particle in position Hilbart space after $t$ steps of the walk is given by, 
\begin{equation}
\ket{\Psi(t)} = \hat{W}^t \bigg[\ket{\psi}_c \otimes \ket{x=0} \bigg ] =  \sum_{x=-t}^{t} \begin{bmatrix} \psi^{\uparrow}_{x, t} \\
 \psi^{\downarrow}_{x, t} \end{bmatrix}.
\end{equation}
The probability of finding the particle at position and time $(x, t)$ is
\begin{align}
P(x, t) = \norm{\psi^{\uparrow}_{x, t}}^2 + \norm{\psi^{\downarrow}_{x, t}}^2 .
\end{align}
Figure. \ref{fig1} shows the probability distribution after 100 steps of a DTQW using a Hadamard coin, where the initial position state is $\ket{0}$.

\section{One-Dimensional Quantum Walk using Three-State Coin in Ternary Quantum System}

 Usually DTQW on the line have two directions to move, right and left. But lazy quantum walks have three choices, right, left and stay put.  In this section, we define the mathematical formalism for lazy quantum walks on one-dimensional line in ternary system.

Dynamics of the lazy quantum walk are defined on the combination of particle (coin) and position Hilbert
space as in DTQW, $\mathcal{H} =\mathcal{H}_c \otimes  \mathcal{H}_p$. A particle with internal states, $
\mathcal{H}_c  = \mbox{span} \{\ket{\uparrow}, \ket{.},  \ket{\downarrow} \}$ ($\uparrow$ for right, $.$ for stay put and $\downarrow$ for left) and  a one-dimensional position Hilbert space is $\mathcal{H}_p  = \mbox{span}
\{\ket{x}\}$, where $x  \in \mathbb{Z}$ represents the labels on the position states in ternary system.

Let $|x,\alpha\rangle$ be a basis state, where $x\in \mathcal{Z}$ represents the position of the particle and $\alpha\in \{\uparrow, ., \downarrow \}$ represents the coin state. The evolution of the whole system at each step of the walk can be described by the unitary operator, denoted by $\hat{U}$,

\begin{equation}
    \hat{U}=\hat{S}({\mathcal{I}}\otimes \hat{C}),
\end{equation}where shift operator($\hat{S}$) is defined by
\begin{equation}\label{22}
  \resizebox{0.99\hsize}{!}{%
  $\hat{S}=\sum_{x\in\mathbb{Z}} \bigg (\ket{\uparrow}\bra{\uparrow}
\otimes   \ket{x-1}\bra{x}+\ket{.}\bra{.}
\otimes   \ket{x}\bra{x}+\ket{\downarrow}\bra{\downarrow} \otimes \ket{x+1}\bra{x}\bigg ).$}%
\end{equation}

\begin{figure}[h!]
\centering
\includegraphics[width=9cm, scale=1]{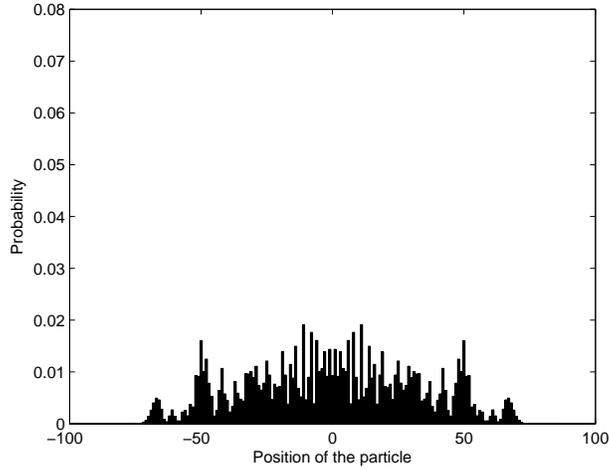}
\captionof{figure}{Lazy quantum walks on a line using three-state DFT coin in ternary system, after 100 steps}
\label{fig2}
\end{figure}

The shift operator at time $t$, translates the
position conditioned on the internal state of the particle. During each step of the LQW, the particle remains at
the same position and also moves to left and right. $I$ is the identity matrix, which operates in $\mathcal{H}_{p}$, while $\hat{C}$ is the coin operation. In this paper, we consider two kinds of coin operators. The first kind is the DFT (Discrete Fourier Transform) coin operator
\begin{equation}
    C=\frac{1}{\sqrt{3}}\left(
          \begin{array}{ccc}
            1 & 1 & 1 \\
            1 & e^{\frac{2\pi i}{3}} & e^{\frac{4\pi i}{3}} \\
            1 & e^{\frac{4\pi i}{3}} & e^{\frac{2\pi i}{3}} \\
          \end{array}
        \right).
\end{equation}
Figure. \ref{fig2} shows the probability distribution after 100 steps of a LQW using three-state DFT coin, where the initial position state is $\ket{0}$ in ternary system.

Besides this coin operator, there are other kinds of $3\times3$ coin operators.

\begin{equation}\label{24}
    G(\rho)=\left(
          \begin{array}{ccc}
            -\rho^2 & \rho \sqrt{2-2\rho^2} & 1-\rho^2 \\
            \rho \sqrt{2-2\rho^2} & 2 \rho^2-1 & \rho \sqrt{2-2\rho^2} \\
            1-\rho^2 & \rho \sqrt{2-2\rho^2} & -\rho^2 \\
          \end{array}
        \right)
\end{equation}
with the coin parameter $\rho\in (0,1)$. This coin operator is equal to the Grover operator when $\rho=\sqrt{\frac{1}{3}}$ (equation. 10).

\begin{equation}
    C=\frac{1}{3}\left(
          \begin{array}{ccc}
            -1 & 2 & 2 \\
            2 & -1 & 2 \\
            2 & 2 & -1 \\
          \end{array}
        \right).
\end{equation}

\subsubsection{Lackadaisical Quantum Walk}

Quantum analogous of special variant of lazy quantum walk, where the walker has some probability of staying put, is known as lackadaisical quantum walk. In lackadaisical quantum walk, the coin degree of freedom is three-dimensional, i.e. $i \in \{\uparrow, \downarrow, . \}$. The flip-flop transformation conditioned on the $\ket{.}$ coin state is

\begin{center}
    $S(\ket{i,j} \otimes \ket{.}) = \ket{i,j} \otimes \ket{.}$
\end{center}

If $l$ self-loops are allowed, then the Coin operator will be $D = 2\ket{s_D}\bra{s_D} - I_3$, where 

\begin{center}
    $\ket{s_D} = \frac{1}{\sqrt{2+l}}(\ket{\uparrow} + \ket{\downarrow}  + \sqrt{l}\ket{.})$
\end{center}

In \cite{Wong_2018}, Wong showed that using lackadaisical quantum walk with $l = \frac{4}{N}$, the success probability of finding a marked state in an $\sqrt{N} \times \sqrt{N}$ grid becomes close to 1 in $\mathcal{O}(\sqrt{NlogN})$ time, which is better than the DTQW.

\section{ Quantum Circuit for Implementing the One-Dimensional Lazy Quantum Walk in Ternary Quantum System}

In this section, we present logical realization of quantum circuits to implement one-dimensional quantum walk using three state coin(LQW) in ternary quantum system. To implement any ternary quantum circuit, we require some dedicated ternary gates which are described in the following subsection.

\subsection{Ternary Gates}
This section gives a brief description of the ternary gates that are required for the circuit synthesis proposed in this paper. 
\subsubsection{Ternary Shift Gates}
In ternary logic, there are five unitary 1-qutrit gates namely, Z(+1), Z(+2), Z(01),Z(12), Z(02)\cite{Muthukrishnan_2000}. These gates were proposed in \cite{Muthukrishnan_2000} where they were realized using ion trap model. Each of these gates can be represented using a $3X3$ unitary matrix. The truth tables for each of these gates are given in Table \ref{tab1}. Fig. \ref{fig3} shows the matrix representations of these gates.

\begin{table}[h]
\centering
\caption{Truth Table of 1-qutrit Z gates}
{%
\begin{tabular}{|c|c|c|c|c|c|}
\hline
A&$Z_{+1}$&$Z_{+2}$&$Z_{01}$&$Z_{12}$&$Z_{02}$\\\hline
0&1&2&1&0&2\\\hline
1&2&0&0&2&1\\\hline
2&0&1&2&1&0\\\hline
\end{tabular}}
\label{tab1}
\end{table}

\begin{figure}[h]
\begin{minipage}[b]{0.15\linewidth}
\centering
$Z_{01}=
\begin{bmatrix}
    0 & 1 & 0 \\
    1 & 0 & 0\\
    0 & 0 & 1
\end{bmatrix}$
\end{minipage}\hfill
\begin{minipage}[b]{0.15\linewidth}
\centering
$Z_{02}=
\begin{bmatrix}
    0 & 0 & 1 \\
    0 & 1 & 0\\
    1 & 0 & 0
\end{bmatrix}$
\end{minipage}\hfill
\begin{minipage}[b]{0.15\linewidth}
\centering
$Z_{12}=
\begin{bmatrix}
    1 & 0 & 0 \\
    0 & 0 & 1\\
    0 & 1 & 0
\end{bmatrix}$
\end{minipage}\hfill
\begin{minipage}[b]{0.15\linewidth}
\centering
$Z_{+1}=
\begin{bmatrix}
    0 & 0 & 1 \\
    1 & 0 & 0\\
    0 & 1 & 0
\end{bmatrix}$
\end{minipage}\hfill
\begin{minipage}[b]{0.15\linewidth}
\centering
$Z_{+2}=
\begin{bmatrix}
    0 & 1 & 0 \\
    0 & 0 & 1\\
    1 & 0 & 0
\end{bmatrix}$
\end{minipage}
\caption{Matrix Representation of 1-qutrit Z gates}
\label{fig3}
\end{figure}

\subsubsection{Ternary Muthukrishnan-Stroud Gate}
\begin{figure}[h]
\centering
\includegraphics[width=8cm, scale=1]{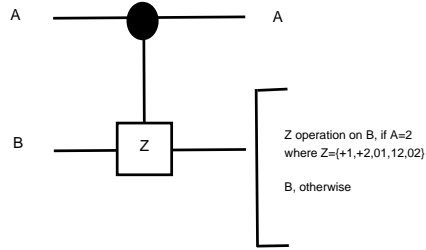}
\captionof{figure}{Circuit Representation of 2-qutrit M-S gate}
\label{fig4}
\end{figure}

Muthukrishnan-Stroud Gate \cite{Muthukrishnan_2000} is a 2-qutrit gate defined as follows:
\[
  M-S_{Z}(A,B) =
  \begin{cases}
                                   \text{apply Z gate to B} & \text{if $A=2$} \\
                                   \text{B} & \text{otherwise} \\
  \end{cases}
\]
where $A$ is the control, $B$ is the target and $Z$ can be any of the five 1-qutrit gates defined above,i.e., $Z=\{+1,+2,01,12,02\}$. The truth tables for all possible 2 qutrit M-S gates are given in Table \ref{tab2}. The general circuit for an M-S gate is given in Fig. \ref{fig4}.\\


\begin{table}[h]
 \caption{Truth Table of 2-qutrit M-S gates}
\centering
\begin{tabular}{|m{2em} | m{2em}| m{2em}  |m{2em} | m{2em}| m{2em}  |  m{2em}|}
\hline
A&B&$M-S_{+1}$&$M-S_{+2}$&$M-S_{01}$&$M-S_{12}$&$M-S_{02}$\\\hline
0&0&0&0&0&0&0\\\hline
0&1&1&1&1&1&1\\\hline
0&2&2&2&2&2&2\\\hline
1&0&0&0&0&0&0\\\hline
1&1&1&1&1&1&1\\\hline
1&2&2&2&2&2&2\\\hline
2&0&1&2&1&0&2\\\hline
2&1&2&0&0&2&1\\\hline
2&2&0&1&2&1&0\\\hline
\end{tabular}
   
    \label{tab2}
\end{table}

\begin{figure}[!h]
\centering
\includegraphics[width=6cm, scale=1]{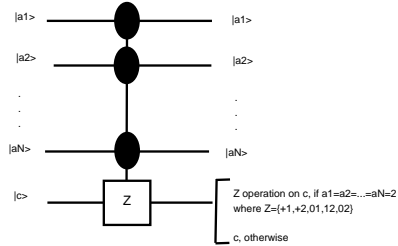}
\caption{Circuit Representation of multi-controlled M-S gate}
\label{fig3a}
\end{figure}

The multi-controlled version of the M-S gates is defined as follows:
 \begin{equation}
\resizebox{.91\hsize}{!}{$ M-S_{Z}(a_1,a_2,\ldots,a_n,c) =\\
  \begin{cases}
                                   \text{apply Z gate to c} & \text{if $a_1=a_2=\ldots=a_n=2$} \\
                                   \text{c} & \text{otherwise} \\
  \end{cases}$}
 \end{equation}
where $a_1,a_2,\ldots,a_n$ are the controlling inputs, $c$ is the target and $Z$ can be any of the five 1-qutrit gate defined above,i.e., $Z=\{+1,+2,01,12,02\}$. The general circuit for a multi-controlled M-S gate is given in Fig. \ref{fig3a}.

\subsection{Implementation}
To implement a LQW in one dimensional position Hilbert space of size $3^q$, $(q+1)$ qutrits
are required, one qutrit to represent the particle's internal state (coin qutrit) and $q$- qutrits to represent the position. The coin operation can be implemented by applying a single qutrit rotation gate on the coin qutrit, and the position shift operation is implemented subsequently with the help of multi-qutrit gates where the coin qutrit acts as the control. Quantum circuits for implementing LQW depends on how the position space is represented. Example circuits for a four qutrit system are given in this section.

For $q=4$ the number of steps of LQW that can be implemented is $\lfloor3^{q-1}/2\rfloor = 13$. We choose the position state mapping given in Table. \ref{tab3}, \ref{tab4}, \ref{tab5} with a fixed initial position state $\ket{000}$. We denote the initial state as $\ket{x = 0} \equiv \ket{000}$ in Table. \ref{tab3}. After first step of LQW, if the coin state is $\ket{2}$, particle moves to the right, $\ket{x = 1} \equiv \ket{002}$, if the coin state is $\ket{1}$, particle moves to the left, $\ket{x = -1} \equiv \ket{001}$, and if the coin state is $\ket{0}$, particle stays at the initial state, $\ket{x = 0} \equiv \ket{000}$ as shown in Table. \ref{tab3}. Fixing the initial state of the walker helps in reducing the gate count in the quantum circuit and hence reduces the overall error. For example, if the initial state is not fixed to $\ket{000}$ then at first we have to bring the initial state to $\ket{000}$ with the help of 1-qutrit M-S gates.

\begin{table}[H]
\centering
\caption{Position state mapping with the multi-qutrits states for quantum circuits presented in Figure. \ref{fig6}}
\begin{tabular}{|m{11.5em} | m{11.5em}  | }
\hline
~~~$\ket{x = 0} \equiv \ket{000}$  &   \\
\hline
~~~$\ket{x = 1} \equiv \ket{002}$ & ~~~$\ket{x = -1} \equiv \ket{001}$\\
\hline 
\end{tabular}
\label{tab3}
\end{table}

\begin{figure}[h]
\centering
\includegraphics[width=5cm, scale=1]{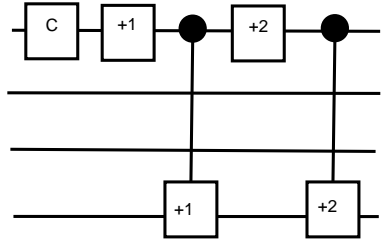}
\captionof{figure}{Quantum circuit for the first step of LQW on four qutrits as given in Table. \ref{tab3}}
\label{fig6}
\end{figure}

\begin{table}[H]
\centering
\caption{Position state mapping with the multi-qutrits states for quantum circuits presented in Figure. \ref{fig7}}
\begin{tabular}{|m{11.5em} | m{11.5em}  | }

\hline
~~~$\ket{x = 2} \equiv \ket{021}$  &  ~~~$\ket{x = -2} \equiv \ket{012}$ \\
\hline
~~~$\ket{x = 3} \equiv \ket{020}$ & ~~~$\ket{x = -3} \equiv \ket{010}$\\
\hline
~~~$\ket{x = 4} \equiv \ket{022}$  & ~~~$\ket{x = -4} \equiv \ket{011}$\\
\hline
\end{tabular}
\label{tab4}
\end{table}

\begin{figure}[h]
\centering
\includegraphics[width=8cm, scale=1]{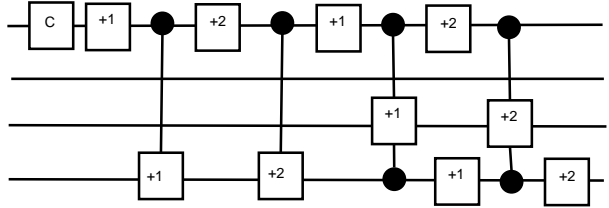}
\captionof{figure}{Quantum circuit for the second-fourth steps of LQW on four qutrits as shown in Table. \ref{tab4}}
\label{fig7}
\end{figure}

\begin{table}[H]
\centering
\caption{Position state mapping with the multi-qutrits states for quantum circuits presented in Figure. \ref{fig8}}
\begin{tabular}{|m{11.5em} | m{11.5em}  | }

\hline
~~~$\ket{x = 5} \equiv \ket{211}$ &  ~~~$\ket{x = -5} \equiv \ket{122}$\\
\hline 
~~~$\ket{x = 6} \equiv \ket{210}$ & ~~~$\ket{x = -6} \equiv \ket{120}$ \\
\hline
~~~$\ket{x = 7} \equiv \ket{212}$ & ~~~$\ket{x = -7} \equiv \ket{121}$ \\
\hline
~~~$\ket{x = 8} \equiv \ket{201}$ & ~~~$\ket{x = -8} \equiv \ket{102}$ \\
\hline
~~~$\ket{x = 9} \equiv \ket{200}$ & ~~~$\ket{x = -9} \equiv \ket{100}$ \\
\hline
~~~$\ket{x = 10} \equiv \ket{202}$ & ~~~$\ket{x = -10} \equiv \ket{101}$ \\
\hline
~~~$\ket{x = 11} \equiv \ket{221}$ & ~~~$\ket{x = -11} \equiv \ket{112}$ \\
\hline
~~~$\ket{x = 12} \equiv \ket{220}$ & ~~~$\ket{x = -12} \equiv \ket{110}$ \\
\hline
~~~$\ket{x = 13} \equiv \ket{222}$ & ~~~$\ket{x = -13} \equiv \ket{111}$ \\
\hline
\end{tabular}
\label{tab5}
\end{table}

\begin{figure}[h]
\centering
\includegraphics[width=8.5cm, scale=1]{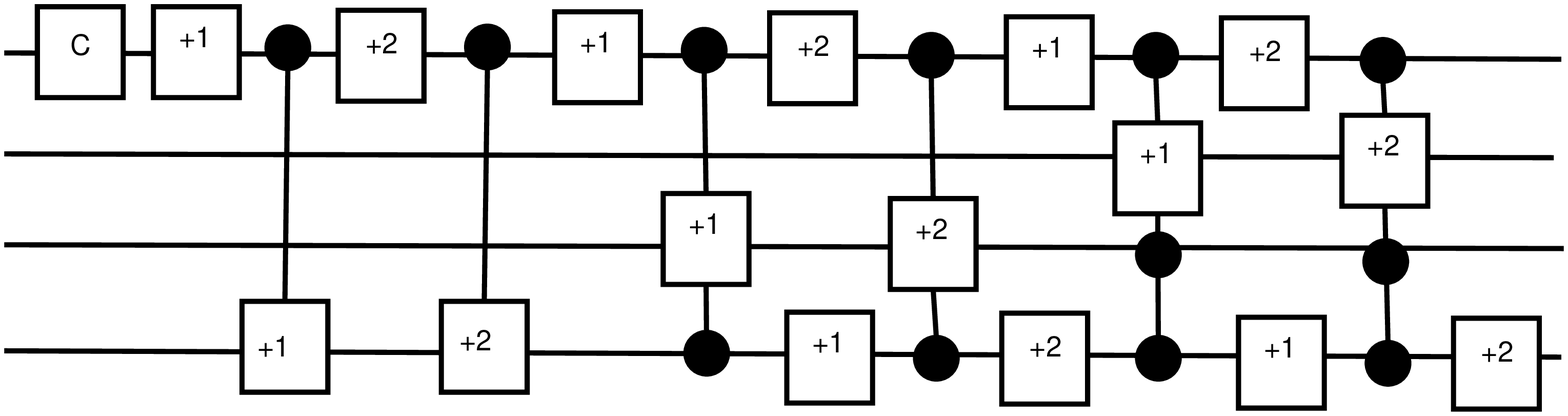}
\captionof{figure}{Quantum circuit for the fifth-thirteenth steps of LQW on four qutrits as shown in Table. \ref{tab5}}
\label{fig8}
\end{figure}

\begin{figure*}[ht!]
\centering
\includegraphics[width=17cm, scale=1]{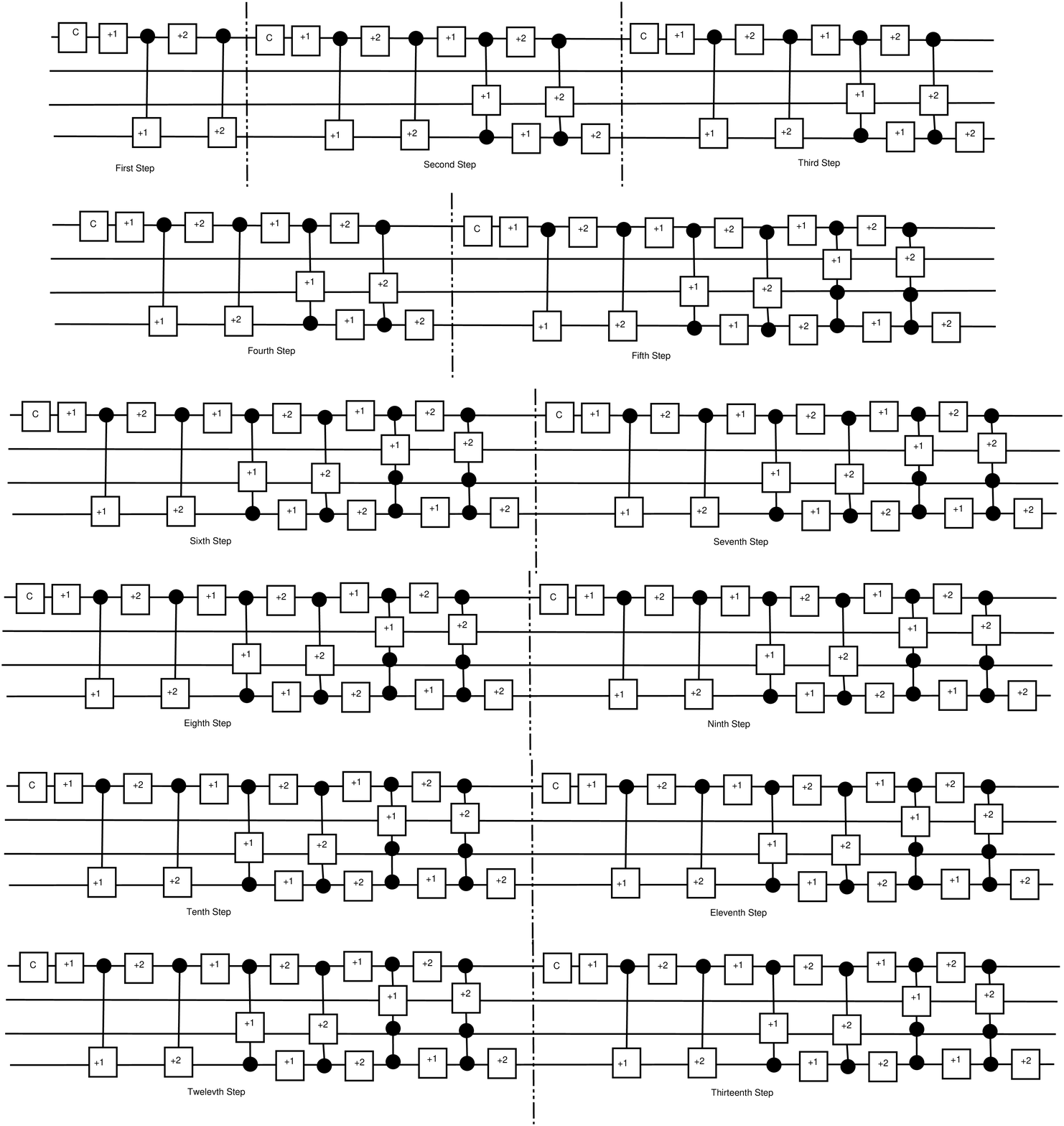}
\captionof{figure}{Quantum circuit for the first thirteen steps of the LQW on a four qutrit ternary system with a fixed initial state $\ket{\uparrow} \otimes \ket{x=0} \equiv \ket{\uparrow} \otimes \ket{000}$. The position state mapping is shown in table\,\ref{tab3}-\ref{tab5}}
\label{fig9}
\end{figure*}

\begin{table}[h]
\centering
\caption{An alternative position state mapping onto the multi-qutrits states for quantum circuits presented in Figure. \ref{fig10}}
\begin{tabular}{|m{11em} | m{11em}  | }

\hline
~~~$\ket{x = 0} \equiv \ket{000}$  &   \\
\hline
~~~$\ket{x = 1} \equiv \ket{001}$ & ~~~$\ket{x = -1} \equiv \ket{002}$\\
\hline 
~~~$\ket{x = 2} \equiv \ket{012}$  &  ~~~$\ket{x = -2} \equiv \ket{021}$ \\
\hline
~~~$\ket{x = 3} \equiv \ket{010}$ & ~~~$\ket{x = -3} \equiv \ket{020}$\\
\hline
~~~$\ket{x = 4} \equiv \ket{011}$  & ~~~$\ket{x = -4} \equiv \ket{022}$\\
\hline
~~~$\ket{x = 5} \equiv \ket{122}$ &  ~~~$\ket{x = -5} \equiv \ket{211}$\\
\hline 
~~~$\ket{x = 6} \equiv \ket{120}$ & ~~~$\ket{x = -6} \equiv \ket{210}$ \\
\hline
~~~$\ket{x = 7} \equiv \ket{121}$ & ~~~$\ket{x = -7} \equiv \ket{212}$ \\
\hline
~~~$\ket{x = 8} \equiv \ket{102}$ & ~~~$\ket{x = -8} \equiv \ket{201}$ \\
\hline
~~~$\ket{x = 9} \equiv \ket{100}$ & ~~~$\ket{x = -9} \equiv \ket{200}$ \\
\hline
~~~$\ket{x = 10} \equiv \ket{101}$ & ~~~$\ket{x = -10} \equiv \ket{202}$ \\
\hline
~~~$\ket{x = 11} \equiv \ket{112}$ & ~~~$\ket{x = -11} \equiv \ket{221}$ \\
\hline
~~~$\ket{x = 12} \equiv \ket{110}$ & ~~~$\ket{x = -12} \equiv \ket{220}$ \\
\hline
~~~$\ket{x = 13} \equiv \ket{111}$ & ~~~$\ket{x = -13} \equiv \ket{222}$ \\
\hline
\end{tabular}
\label{tab7}
\end{table}

\begin{figure*}[ht!]
\centering
\includegraphics[width=12cm, scale=1]{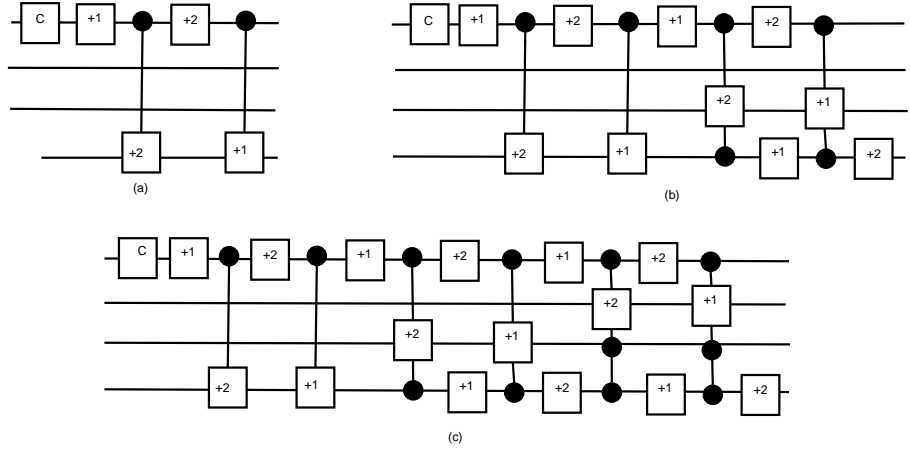}
\captionof{figure}{(a)Quantum circuit for the first steps of LQW on four qutrits as shown in Table. \ref{tab7}; (b)Quantum circuit for the second-fourth steps of LQW on four qutrits as shown in Table. \ref{tab7} (c)Quantum circuit for the fifth-thirteenth steps of LQW on four qutrits as shown in Table. \ref{tab7}}
\label{fig10}
\end{figure*}

\begin{table} [ht!]
\centering
\caption{Number of steps and maximum number of control qutrits required to control a target qutrit in the LQW for a ternary system of upto $n$ qutrits using  a circuit similar to the one presented in Figure. \ref{fig9}}
\begin{tabular}{|m{6em}| m{6em} | m{7.5em} |}
\hline
No. of qutrits & No. of steps & Max. No. of controls in M$-$S gates\\
\hline
 2 & 1 & 1 \\
\hline
 3 & 4 & 2 \\
\hline 
 4 & 13 & 3\\
\hline
 5 & 40 & 4 \\
\hline
$n+1$ & $\lfloor 3^n/2 \rfloor$ & n\\
\hline
\end{tabular}
\label{tab8}
\end{table}

After each step of the LQW, two new position states have to be considered along with stay put. In Table. \ref{tab3}, \ref{tab4}, \ref{tab5}, we show that the mapping of these new position states onto the multi-qutrit states is in such a way that optimal number of gates are used to implement the shift operation, we consider the nearest neighbour position space so as to make the circuit efficient. In Figure. \ref{fig6}, \ref{fig7}, \ref{fig8}, Quantum circuits are developed for mapping of position states, which are shown in Table. \ref{tab3}, \ref{tab4}, \ref{tab5} respectively. Quantum circuit for first thirteen steps of the LQW on line by considering the position state is mapping shown in Table. \ref{tab3}-\ref{tab5} on a four qutrit ternary system with a fixed initial state $\ket{\uparrow} \otimes \ket{x=0} \equiv \ket{\uparrow} \otimes \ket{000}$, they have been illustrated in Figure. \ref{fig9}. The output of the quantum circuit shown in Figure. \ref{fig8} for first three steps of LQW using three-state DFT coin is illustrated in Table. \ref{tab9}.

\begin{figure*}[ht!]
\centering
\includegraphics[width=16cm, scale=1]{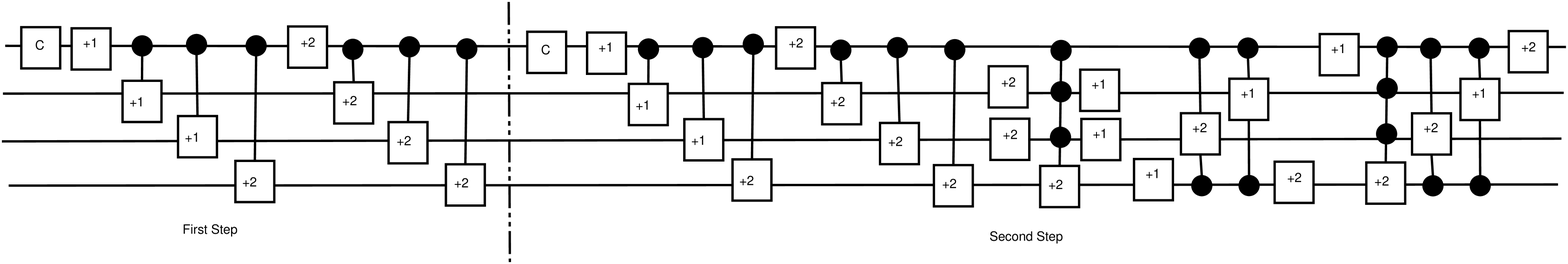}
\captionof{figure}{Quantum circuit for first two steps of the LQW on a four qutrit ternary system with a fixed initial state $\ket{\uparrow} \otimes \ket{x=0} \equiv \ket{\uparrow} \otimes \ket{000}$. The position state mapping is shown in Table.\,\ref{example}}
\label{examplepic}
\end{figure*}

An alternative quantum circuit is shown in Figure \ref{fig10} for different mapping choice of position states (As we will always have two alternatives nearest neighbour position spaces due to  three orthonormal basis states in ternary system) onto multi-qutrits states is shown in the Table. \ref{tab7}. These two mapping choices are the only appropriate mapping of qutrit states with the nearest neighbour position space, which makes the circuit not only efficient but also generalized for $n$ qutrit systems. Apart from these mapping choices, any of naive mapping choices of the position states onto the qutrit states will lead into an inefficient quantum circuit with higher number of quantum gates as they don't follow nearest neighbour logic. One such example is given in Table. \ref{example} and Figure. \ref{examplepic}. In this example, due to the configuration of mapped position state, only two steps of LQW can be performed, which is shown in Figure. \ref{examplepic} whereas, using our position state mapping approach, $13$ steps of LQW can be realized in the same system. This naive mapping choices of the position states based circuit consists of many additional ternary M-S gates compared to the nearest neighbour position space based circuits shown in Figure. \ref{fig9}. The next subsection will discuss about the generalization of quantum circuit for the LQW on a $n$ qutrit ternary system.

\begin{table}[h]
\centering
\caption{An example of mapping of position state onto the multi-qutrits states for quantum circuits presented in Figure. \ref{examplepic}}
\begin{tabular}{|m{11em} | m{11em}  | }

\hline
~~~$\ket{x = 0} \equiv \ket{000}$  &   \\
\hline
~~~$\ket{x = 1} \equiv \ket{222}$ & ~~~$\ket{x = -1} \equiv \ket{112}$\\
\hline 
~~~$\ket{x = 2} \equiv \ket{021}$  &  ~~~$\ket{x = -2} \equiv \ket{012}$ \\
\hline
~~~$\ket{x = 3} \equiv \ket{212}$ & ~~~$\ket{x = -3} \equiv \ket{211}$\\
\hline
\end{tabular}
\label{example}
\end{table}

\begin{table*}[ht]
\centering
\caption{Output after each step(first three steps) of LQW using three-state DFT coin and output of quantum circuit shown in Figure. \ref{fig8}}
\begin{tabular}{ | m{4em} | m{20em} | m{20em} | 
} 
\hline
$~~~$Steps &  Lazy quantum walk output & Circuit (Figure. \ref{fig8}) output\\
\hline
$~~~~$0   &     $\ket{0} \otimes \ket{x= 0} $      &      $ \ket{0} \otimes \ket{000} $ \\ 
\hline 
$~~~~$1   &     $ (\ket{0} \otimes \ket{x= 0} +  \ket{1} \otimes \ket{x=-1} +  \ket{2} \otimes \ket{x=1})/\sqrt{3}$     & $(\ket{0} \otimes \ket{000} + \ket{1} \otimes \ket{001} + \ket{2} \otimes \ket{002})/\sqrt{3}$ \\ 
\hline 
$~~~~$2    &   $ (\ket{0} \otimes \ket{x= 0} +  \ket{1} \otimes \ket{x=-1} +  \ket{2} \otimes \ket{x=1} +  
\ket{0} \otimes \ket{x= -1} +  e^{\frac{2\pi i}{3}} * \ket{1} \otimes \ket{x=-2} +  e^{\frac{4\pi i}{3}} * \ket{2} \otimes \ket{x=0}  + 
\ket{0} \otimes \ket{x= 1} + e^{\frac{4\pi i}{3}} * \ket{1} \otimes \ket{x=0} + e^{\frac{2\pi i}{3}} * \ket{2} \otimes \ket{x=2})/3$    &   $(\ket{0} \otimes \ket{000} + \ket{1} \otimes \ket{001} + \ket{2} \otimes \ket{002} + 
\ket{0} \otimes \ket{001} + e^{\frac{2\pi i}{3}} * \ket{1} \otimes \ket{012} + e^{\frac{4\pi i}{3}} * \ket{2} \otimes \ket{000} + \ket{0} \otimes \ket{002} + e^{\frac{4\pi i}{3}} * \ket{1} \otimes \ket{000} + e^{\frac{4\pi i}{3}} * \ket{2} \otimes \ket{021})/3$   \\ 
\hline
$~~~~$3   &  $ (\ket{0} \otimes \ket{x= 0} +  \ket{1} \otimes \ket{x=-1} +  \ket{2} \otimes \ket{x=1} +  
\ket{0} \otimes \ket{x= -1} +  e^{\frac{2\pi i}{3}} * \ket{1} \otimes \ket{x=-2} +  e^{\frac{4\pi i}{3}} * \ket{2} \otimes \ket{x=0}  + 
\ket{0} \otimes \ket{x= 1} + e^{\frac{4\pi i}{3}} * \ket{1} \otimes \ket{x=0} + e^{\frac{2\pi i}{3}} * \ket{2} \otimes \ket{x=2} + \ket{0} \otimes \ket{x= -1} +  \ket{1} \otimes \ket{x=-2} +  \ket{2} \otimes \ket{x=0} +  
e^{\frac{2\pi i}{3}} *
(\ket{0} \otimes \ket{x= -2} +  e^{\frac{2\pi i}{3}} * \ket{1} \otimes \ket{x=-3} +  e^{\frac{4\pi i}{3}} * \ket{2} \otimes \ket{x=-1})  + e^{\frac{4\pi i}{3}} *
(\ket{0} \otimes \ket{x= 0} + e^{\frac{4\pi i}{3}} * \ket{1} \otimes \ket{x=-1} + e^{\frac{2\pi i}{3}} * \ket{2} \otimes \ket{x=1}) + 
\ket{0} \otimes \ket{x= 1} +  \ket{1} \otimes \ket{x=0} +  \ket{2} \otimes \ket{x=2} + e^{\frac{4\pi i}{3}} *  
(\ket{0} \otimes \ket{x= 0} +  e^{\frac{2\pi i}{3}} * \ket{1} \otimes \ket{x=-1} +  e^{\frac{4\pi i}{3}} * \ket{2} \otimes \ket{x=1})  + e^{\frac{2\pi i}{3}} * 
(\ket{0} \otimes \ket{x= 2} + e^{\frac{4\pi i}{3}} * \ket{1} \otimes \ket{x=1} + e^{\frac{2\pi i}{3}} * \ket{2} \otimes \ket{x=3}))/3\sqrt{3}$  &    $(\ket{0} \otimes \ket{000} + \ket{1} \otimes \ket{001} + \ket{2} \otimes \ket{002} + 
\ket{0} \otimes \ket{001} + e^{\frac{2\pi i}{3}} * \ket{1} \otimes \ket{012} + e^{\frac{4\pi i}{3}} * \ket{2} \otimes \ket{000} + \ket{0} \otimes \ket{002} + e^{\frac{4\pi i}{3}} * \ket{1} \otimes \ket{000} + e^{\frac{4\pi i}{3}} * \ket{2} \otimes \ket{021} + \ket{0} \otimes \ket{001} + \ket{1} \otimes \ket{012} + \ket{2} \otimes \ket{000} + e^{\frac{2\pi i}{3}} *(
\ket{0} \otimes \ket{012} + e^{\frac{2\pi i}{3}} * \ket{1} \otimes \ket{010} + e^{\frac{4\pi i}{3}} * \ket{2} \otimes \ket{001}) + e^{\frac{4\pi i}{3}} *( \ket{0} \otimes \ket{000} + e^{\frac{4\pi i}{3}} * \ket{1} \otimes \ket{001} + e^{\frac{4\pi i}{3}} * \ket{2} \otimes \ket{002}) + \ket{0} \otimes \ket{000} + \ket{1} \otimes \ket{001} + \ket{2} \otimes \ket{002} + e^{\frac{4\pi i}{3}} *
(\ket{0} \otimes \ket{000} + e^{\frac{2\pi i}{3}} * \ket{1} \otimes \ket{001} + e^{\frac{4\pi i}{3}} * \ket{2} \otimes \ket{002}) + e^{\frac{2\pi i}{3}} * (\ket{0} \otimes \ket{021} + e^{\frac{4\pi i}{3}} * \ket{1} \otimes \ket{002} + e^{\frac{4\pi i}{3}} * \ket{2} \otimes \ket{02}))/3\sqrt{3}$\\ 
\hline

\end{tabular}
\label{tab9}
\end{table*}

\subsection{Generalization of Quantum Circuit for Implementing One-Dimensional Lazy Quantum Walk in Ternary System }

\begin{figure*}[ht!]
\centering
\includegraphics[width=16cm, scale=1]{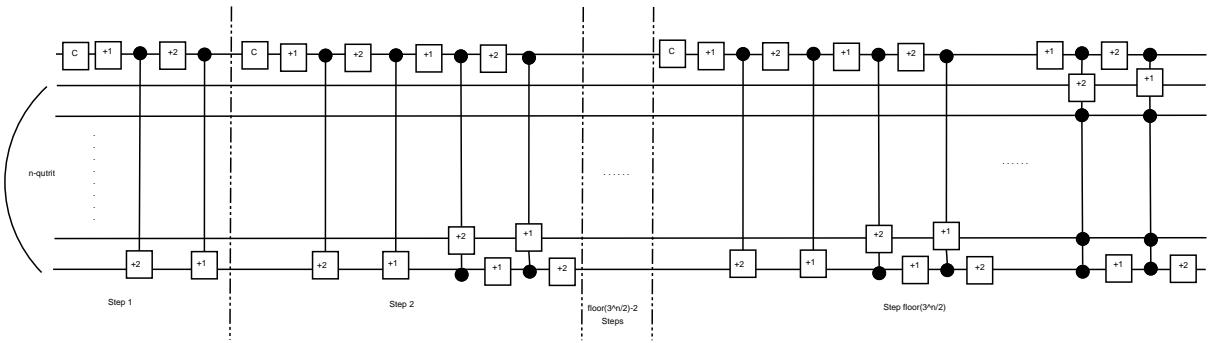}
\captionof{figure}{Generalization of Quantum circuit for $\lfloor 3^n/2 \rfloor$ steps of the LQW on a $n$ qutrit ternary system with a fixed initial state $\ket{\uparrow} \otimes \ket{x=0} \equiv \ket{\uparrow} \otimes \ket{(n-1)\; times\; 0's}$.}
\label{fig11}
\end{figure*}

The proposed circuits can be scaled to implement more steps on a larger ternary system with the help of higher controlled M-S gates. As shown in Table \ref{tab8}, using $n$-qutrit system, implementation of $\lfloor3^{n}/2\rfloor$-steps of a LQW can be performed. In other words, to implement $n$-step of LQW, at least $((\lfloor\log_{3}n\rfloor+1) + 1)$ qutrits are required. Figure. \ref{fig11} shows the generalized quantum circuit for $\lfloor 3^n/2 \rfloor$ steps of the LQW on a $n$ qutrit ternary system with a fixed initial state $\ket{\uparrow} \otimes \ket{x=0} \equiv \ket{\uparrow} \otimes \ket{(n-1)\; times\; 0's}$. As shown in Table \ref{tab6}, if the position state mapping onto multi-qutrits states can be scaled up to $\lfloor3^{n}/2\rfloor$-steps, the generalized quantum circuit on $n$-qutrit ternary system shown in Figure \ref{fig11} will be required to implement LQW. As shown in our proposed circuit, for a $n$-qutrit system after every $3^{q-1}$ (where $q$ is the number of qutrits and $q$ ranges from 1 to $n$) steps, two new gates are added to realize the new position states along with the previous set of gates (when $q>1$) due to stay put. As discussed in previous subsection an alternative way of position state mapping onto the multi-qutrit states exists, which can also be generalized similarly.

\begin{table*}[ht!]
\centering
\caption{Position state mapping onto the multi-qutrits states for quantum circuits presented in Figure. \ref{fig11}.}
\begin{tabular}{|m{3.5em} | m{3.5em}| m{3.5em} | m{3.5em}|  m{3.5em} | m{3.5em}|  m{3.5em} | m{3.5em}|  m{3.5em} | m{3.5em}|  m{3.5em} |}

\hline
$\ket{x = 5}$&$\ket{x = 4}$&$\ket{x = 3}$&$\ket{x = 2}$&$\ket{x = 1}$&$\ket{x = 0} $  &$\ket{x = -1}$&$\ket{x = -2}$&$\ket{x = -3}$&$\ket{x = -4}$&$\ket{x = -5}$  \\
\hline
&&&&&$ \ket{000}$  &&&&&  \\
\hline
&&&&$ \ket{002}$&$ \ket{000}$  &$ \ket{001}$&&&&  \\
\hline
&&&$ \ket{021}$&$ \ket{002}$&$ \ket{000}$  &$ \ket{001}$&$ \ket{012}$&&&  \\
\hline
&&$ \ket{020}$&$ \ket{021}$&$ \ket{002}$&$ \ket{000}$ &$ \ket{001}$&$ \ket{012}$&$ \ket{010}$&&  \\
\hline
&$ \ket{022}$&$ \ket{020}$&$ \ket{021}$&$ \ket{002}$&$ \ket{000}$ &$ \ket{001}$&$ \ket{012}$&$ \ket{010}$&$ \ket{011}$&  \\
\hline
$ \ket{211}$&$ \ket{022}$&$ \ket{020}$&$ \ket{021}$&$ \ket{002}$&$ \ket{000}$ &$ \ket{001}$&$ \ket{012}$&$ \ket{010}$&$ \ket{011}$& $ \ket{122}$ \\
\hline

\end{tabular}
\label{tab6}
\end{table*}

\section{Conclusion}
In this work, we have defined LQW in ternary quantum system. Further, we have proposed an efficient quantum circuit realization to implement LQW in ternary quantum system using an appropriate logical mapping of the position space on which a walker evolves onto the multi-qutrit states. Later, we also address scalability of the proposed circuit to $n$-qutrit system. We have also verified our proposed circuits through simulation in Matlab. To implement a three-state LQW in two-dimensional position space, the same circuit can be scaled with an appropriate mapping of qutrit states with the nearest neighbour position space in both dimensions by introducing one more coin qutrit into the ternary system. This approach can also be scaled up to $n$-dimensional position space. In such cases, the control over the target or position qutrit increases with the number of coin qutrits. These circuits can also be implemented in any ternary quantum devices.

\section{Acknowledgments}
We gratefully acknowledge fruitful discussion regarding Quantum walk using three-state coin with Prof. C. M. Candrashekar, Insitute of Mathematical Science. We also thankful to Prof. Andris Ambainis, University of Latvia for the discussion on behaviour analysis and algorithmic implementation of Quantum walk. This work has been supported by the grant from CSIR, Govt. of India, Grant No. 09/028(0987)/2016-EMR-I.

\bibliographystyle{unsrt}  

\bibliography{references}

\begin{thebibliography}{10}

\bibitem{nielsen_chuang_2010}
Michael~A. Nielsen and Isaac~L. Chuang.
\newblock {\em Quantum Computation and Quantum Information: 10th Anniversary
  Edition}.
\newblock Cambridge University Press, 2010.

\bibitem{Farhi_1998}
Edward Farhi and Sam Gutmann.
\newblock Quantum computation and decision trees.
\newblock {\em Physical Review A}, 58(2):915–928, Aug 1998.

\bibitem{PhysRevA.48.1687}
Y.~Aharonov, L.~Davidovich, and N.~Zagury.
\newblock Quantum random walks.
\newblock {\em Phys. Rev. A}, 48:1687--1690, Aug 1993.

\bibitem{Shenvi_2003}
Neil Shenvi, Julia Kempe, and K.~Birgitta Whaley.
\newblock Quantum random-walk search algorithm.
\newblock {\em Physical Review A}, 67(5), May 2003.

\bibitem{ambainis2004quantum}
Andris Ambainis.
\newblock Quantum walks and their algorithmic applications, 2004.

\bibitem{aharonov2000quantum}
Dorit Aharonov, Andris Ambainis, Julia Kempe, and Umesh Vazirani.
\newblock Quantum walks on graphs, 2000.

\bibitem{magniez2003quantum}
Frederic Magniez, Miklos Santha, and Mario Szegedy.
\newblock Quantum algorithms for the triangle problem, 2003.

\bibitem{ambainis2003quantum}
Andris Ambainis.
\newblock Quantum walk algorithm for element distinctness, 2003.

\bibitem{Tregenna_2003}
Ben Tregenna, Will Flanagan, Rik Maile, and Viv Kendon.
\newblock Controlling discrete quantum walks: coins and initial states.
\newblock {\em New Journal of Physics}, 5:83–83, Jul 2003.

\bibitem{Childs_2003}
Andrew~M. Childs, Richard Cleve, Enrico Deotto, Edward Farhi, Sam Gutmann, and
  Daniel~A. Spielman.
\newblock Exponential algorithmic speedup by a quantum walk.
\newblock {\em Proceedings of the thirty-fifth ACM symposium on Theory of
  computing - STOC ’03}, 2003.

\bibitem{Kempe03}
J~Kempe.
\newblock Quantum random walks: An introductory overview.
\newblock {\em Contemporary Physics}, 44(4):307–327, Jul 2003.

\bibitem{Venegas_Andraca_2012}
Salvador~Elías Venegas-Andraca.
\newblock Quantum walks: a comprehensive review.
\newblock {\em Quantum Information Processing}, 11(5):1015–1106, Jul 2012.

\bibitem{Kendon_2006}
Vivien~M Kendon.
\newblock A random walk approach to quantum algorithms.
\newblock {\em Philosophical Transactions of the Royal Society A: Mathematical,
  Physical and Engineering Sciences}, 364(1849):3407–3422, Oct 2006.

\bibitem{Childs_2009}
Andrew~M. Childs.
\newblock On the relationship between continuous- and discrete-time quantum
  walk.
\newblock {\em Communications in Mathematical Physics}, 294(2):581–603, Oct
  2009.

\bibitem{Falkner_2014}
Stefan Falkner and Stefan Boettcher.
\newblock Weak limit of the three-state quantum walk on the line.
\newblock {\em Physical Review A}, 90(1), Jul 2014.

\bibitem{machida2014limit}
Takuya Machida.
\newblock Limit theorems of a 3-state quantum walk and its application for
  discrete uniform measures, 2014.

\bibitem{endo2019stationary}
Takako Endo, Takashi Komatsu, Norio Konno, and Tomoyuki Terada.
\newblock Stationary measure for three-state quantum walk, 2019.

\bibitem{kajiwara2019periodicity}
Takeshi Kajiwara, Norio Konno, Shohei Koyama, and Kei Saito.
\newblock Periodicity for the 3-state quantum walk on cycles, 2019.

\bibitem{Li_2015}
Dan Li, Michael~Mc Gettrick, Wei-Wei Zhang, and Ke-Jia Zhang.
\newblock One-dimensional lazy quantum walks and occupancy rate.
\newblock {\em Chinese Physics B}, 24(5):050305, Apr 2015.

\bibitem{Wong_2015}
Thomas~G Wong.
\newblock Grover search with lackadaisical quantum walks.
\newblock {\em Journal of Physics A: Mathematical and Theoretical},
  48(43):435304, Oct 2015.

\bibitem{Wong_2018}
Thomas~G. Wong.
\newblock Faster search by lackadaisical quantum walk.
\newblock {\em Quantum Information Processing}, 17(3), Feb 2018.

\bibitem{saha2018search}
Amit Saha, Ritajit Majumdar, Debasri Saha, Amlan Chakrabarti, and Susmita
  Sur-Kolay.
\newblock Search of clustered marked states with lackadaisical quantum walks,
  2018.

\bibitem{balu2017physical}
Radhakrishnan Balu, Daniel Castillo, and George Siopsis.
\newblock Physical realization of topological quantum walks on ibm-q and
  beyond, 2017.

\bibitem{Yan753}
Zhiguang Yan, Yu-Ran Zhang, Ming Gong, Yulin Wu, Yarui Zheng, Shaowei Li, Can
  Wang, Futian Liang, Jin Lin, Yu~Xu, Cheng Guo, Lihua Sun, Cheng-Zhi Peng,
  Keyu Xia, Hui Deng, Hao Rong, J.~Q. You, Franco Nori, Heng Fan, Xiaobo Zhu,
  and Jian-Wei Pan.
\newblock Strongly correlated quantum walks with a 12-qubit superconducting
  processor.
\newblock {\em Science}, 364(6442):753--756, 2019.

\bibitem{alderete2020quantum}
C.~Huerta Alderete, Shivani Singh, Nhung~H. Nguyen, Daiwei Zhu, Radhakrishnan
  Balu, Christopher Monroe, C.~M. Chandrashekar, and Norbert~M. Linke.
\newblock Quantum walks and dirac cellular automata on a programmable
  trapped-ion quantum computer, 2020.

\bibitem{acasiete2020experimental}
Frank Acasiete, Flavia~P. Agostini, Jalil~Khatibi Moqadam, and Renato Portugal.
\newblock Experimental implementation of quantum walks on ibm quantum
  computers, 2020.

\bibitem{singh2020universal}
Shivani Singh, Cinthia~H. Alderete, Radhakrishnan Balu, Christopher Monroe,
  Norbert~M. Linke, and C.~M. Chandrashekar.
\newblock Universal one-dimensional discrete-time quantum walks and their
  implementation on near term quantum hardware, 2020.

\bibitem{Zhou_2019}
Jia-Qi Zhou, Ling Cai, Qi-Ping Su, and Chui-Ping Yang.
\newblock Protocol of a quantum walk in circuit qed.
\newblock {\em Physical Review A}, 100(1), Jul 2019.

\bibitem{Muthukrishnan_2000}
Ashok Muthukrishnan and C.~R. Stroud.
\newblock Multivalued logic gates for quantum computation.
\newblock {\em Physical Review A}, 62(5), Oct 2000.

\bibitem{Fan_2007}
Yale Fan.
\newblock A generalization of the deutsch-jozsa algorithm to multi-valued
  quantum logic.
\newblock {\em 37th International Symposium on Multiple-Valued Logic
  (ISMVL’07)}, May 2007.

\bibitem{bocharov2015improved}
Alex Bocharov, Shawn~X. Cui, Martin Roetteler, and Krysta~M. Svore.
\newblock Improved quantum ternary arithmetics, 2015.

\bibitem{Bocharov_2017}
Alex Bocharov, Martin Roetteler, and Krysta~M. Svore.
\newblock Factoring with qutrits: Shor’s algorithm on ternary and metaplectic
  quantum architectures.
\newblock {\em Physical Review A}, 96(1), Jul 2017.

\bibitem{6845018}
S.~B. {Mandal}, A.~{Chakrabarti}, and S.~{Sur-Kolay}.
\newblock Synthesis of ternary grover's algorithm.
\newblock In {\em 2014 IEEE 44th International Symposium on Multiple-Valued
  Logic}, pages 184--189, 2014.

\bibitem{Ivanov_2012}
S.~S. Ivanov, H.~S. Tonchev, and N.~V. Vitanov.
\newblock Time-efficient implementation of quantum search with qudits.
\newblock {\em Physical Review A}, 85(6), Jun 2012.

\bibitem{doi:10.1142/S0218126615501212}
Fuyou Fan, Guowu Yang, Gang Yang, and William N.~N. Hung.
\newblock A synthesis method of quantum reversible logic circuit based on
  elementary qutrit quantum logic gates.
\newblock {\em Journal of Circuits, Systems and Computers}, 24(08):1550121,
  2015.

\bibitem{di2011elementary}
Yao-Min Di and Hai-Rui Wei.
\newblock Elementary gates for ternary quantum logic circuit, 2011.

\bibitem{Di_2013}
Yao-Min Di and Hai-Rui Wei.
\newblock Synthesis of multivalued quantum logic circuits by elementary gates.
\newblock {\em Physical Review A}, 87(1), Jan 2013.

\bibitem{blok2020quantum}
M.~S. Blok, V.~V. Ramasesh, T.~Schuster, K.~O'Brien, J.~M. Kreikebaum,
  D.~Dahlen, A.~Morvan, B.~Yoshida, N.~Y. Yao, and I.~Siddiqi.
\newblock Quantum information scrambling in a superconducting qutrit processor,
  2020.

\bibitem{yurtalan2020implementation}
M.~A. Yurtalan, J.~Shi, M.~Kononenko, A.~Lupascu, and S.~Ashhab.
\newblock Implementation of a walsh-hadamard gate in a superconducting qutrit,
  2020.

\end{thebibliography}

\end{document}